# SnakeLines: integrated set of computational pipelines for sequencing reads

Jaroslav Budiš[1,2,3,*], Werner Krampl[1,3,5], Marcel Kucharík[1,3], Rastislav Hekel[1,2,5], Adrián Goga[3,4], Michal Lichvár[1,3], Dávid Smoľak[1,5], Miroslav Böhmer[1,3,5], Andrej Baláž[1,5], František Ďuriš[1,2], Juraj Gazdarica[1,2,3], Katarína Šoltys[3,6], Ján Turňa[2,3,6], Ján Radvánszky[1,3,7], Tomáš Szemes[1,3,6]

[1]Geneton Ltd., 841 04 Bratislava, Slovakia

[2]Slovak Centre of Scientific and Technical Information, 811 04 Bratislava, Slovakia

[3]Comenius University Science Park, 841 04 Bratislava, Slovakia

[4]Department of Computer Science, Faculty of Mathematics, Physics and Informatics, Comenius University, 841 04 Bratislava, Slovakia

[5]Department of Applied Informatics, Faculty of Mathematics, Physics and Informatics, Comenius University, 841 04 Bratislava, Slovakia

[6]Department of Molecular Biology, Faculty of Natural Sciences, Comenius University, 841 04 Bratislava, Slovakia

[7]Institute of Clinical and Translational Research, Biomedical Research Center, Slovak Academy of Sciences, 845 05 Bratislava, Slovakia

# Abstract

**Background:** With the rapid growth of massively parallel sequencing technologies, still more laboratories are utilizing sequenced DNA fragments for genomic analyses. Interpretation of sequencing data is, however, strongly dependent on bioinformatics processing, which is often too demanding for clinicians and researchers without a computational background. Another problem represents the reproducibility of computational analyses across separated computational centers with inconsistent versions of installed libraries and bioinformatics tools.

**Results:** We propose an easily extensible set of computational pipelines, called SnakeLines, for processing sequencing reads; including mapping, assembly, variant calling, viral identification, transcriptomics, metagenomics, and methylation analysis. Individual steps of an analysis, along with methods and their parameters can be readily modified in a single configuration file. Provided pipelines are embedded in virtual environments that ensure isolation of required resources from the host operating system, rapid deployment, and reproducibility of analysis across different Unix-based platforms.

**Conclusion:** SnakeLines is a powerful framework for the automation of bioinformatics analyses, with emphasis on a simple set-up, modifications, extensibility, and reproducibility.

# Background

Massively parallel sequencing (MPS) technologies have revolutionized not only research in molecular biology but also several clinical fields associated with genomic analyses. The rapid increase of genomic data has brought new challenges, mainly in transforming raw sequencing data into results interpretable by researchers and clinicians. Besides computational challenges, operatives must deal with a wide spectrum of available bioinformatics tools that are typically connected in computational pipelines. Development and testing of pipelines usually take a considerable amount of time and the process is prone to errors that are difficult to identify from the output files alone. Another problem is to ensure the reproducibility of the analysis across separated computational centers or distinct platforms with inconsistent software versions (Munafò et al. 2017).

Several systems for the management of pipelines have been described and released to handle complex processing steps associated with MPS data (Leipzig 2017). All of these have stronger and weaker sides. For example, frameworks based on graphical interfaces (Afgan et al. 2018; Wolstencroft et al. 2013) are suitable for researchers without a strong computational background. On the other hand, frameworks based on the command-line interface (CLI) are more flexible, and so are typically preferred by bioinformaticians (Cingolani, Sladek, and Blanchette 2015; H Backman and Girke 2016; Joo et al. 2019). Lately, the CLI-based Snakemake workflow engine (Köster and Rahmann 2012) gained a lot of attention, leading to several bioinformatic pipelines for various domains, such as metagenomics (Piro, Matschkowski, and Renard 2017), variant calling (Cokelaer et al. 2017), transcriptomics (D. Wang 2018; Cornwell et al. 2018; Singer et al. 2018) and ChIP-seq data (Rioualen,

Charbonnier-Khamvongsa, and van Helden 2017). Although these pipelines are usually not limited to a single domain, sequence centers with broader scope may use several of them, and so handle multiple installations with different approaches to external dependencies, configuration, and execution (Table 1).

We propose a set of Snakemake pipelines for a wider spectrum of bioinformatics analyses, called SnakeLines. The framework is designed to be easily extensible and adjustable with a single user-defined configuration file with an emphasis on the rapid deployment of required software and reproducibility of computational analysis. Although the pipelines were primarily developed for paired-end Illumina reads, they can be readily extended with the single-end Illumina and Nanopore specific tools to be applied to a wider set of sequencing technologies.

The open-source code of the proposed methods, together with test data, is freely available for non-commercial users at https://github.com/jbudis/snakelines along with Anaconda repository https://anaconda.org/bioconda/snakelines for rapid set-up and installation. Description of implemented pipelines, as well as instructions for installation, running, and extending the framework, are accessible from the online documentation https://snakelines.readthedocs.io/.

## Material and Methods

SnakeLines pipelines are compiled and executed by the Snakemake workflow engine.

Snakemake rules represent atomic operations of pipelines, such as trimming of reads, mapping to reference sequences, or variant calling. Such rules are defined by mandatory input files, generated output files, and source code of operations that transform the input files into the output files. Dependencies between rules are automatically determined by Snakemake, defining a succession of operations that generates requested output files. The SnakeLines framework extends the Snakemake engine with three main components: 1) the set of configurable rule templates for commonly used bioinformatics tools; 2) the set of custom Python scripts that build up the Snakemake pipeline from the rule templates based on the user-defined configuration file; 3) and the definition of virtual environments with all bioinformatics tools required for the execution of the pipeline (Figure 1).

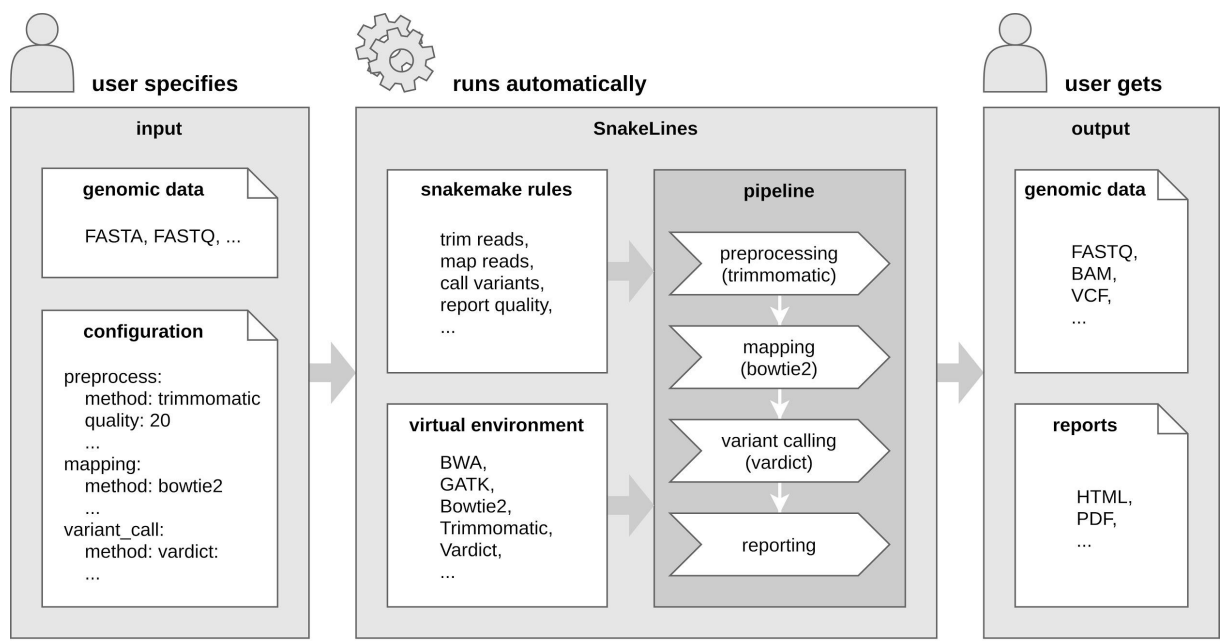

*Figure 1.* *Standard execution of a SnakeLines pipeline. The user supplies the configuration file and genomic data originating from a sequencing run together with the required set of pipeline-specific files, such as a reference genome. Based on the configuration, SnakeLines identifies required Snakemake rules, bioinformatics tools, and their parameters. The exact steps of the computational pipeline are dynamically assembled and automatically executed in*

*virtual environments using the Snakemake workflow engine. The output of the pipeline is a set of generated genomic files and a set of associated quality reports.*

Since SnakeLines pipelines are executed with standard Snakemake calls, users may utilize its generous set of extended features, such as visualization, monitoring, and parallel execution of pipelines that can be distributed over several computational nodes. SnakeLines adds extra functionality that allows simple set-up and modification of a pipeline from the single configuration file; including parameterization of used bioinformatics tools, their replacement with provided or custom alternatives, and omission or addition of requested processing steps. The user has overall information on all operations, as well as executed tools and their parameters. All required tools are automatically set up into isolated virtual environments to avoid common problems with their installation and inconsistent dependencies. SnakeLines uses Conda package repositories since they represent the most extensive source of bioinformatics packages from various language ecosystems used in the field (Grüning et al. 2018). This approach ensures the reproducibility of the analysis across different computational centers since all tools are installed in the same predefined versions.

## Configuration of pipelines

Each SnakeLines pipeline is entirely defined by its configuration file in the YAML format. The user only has to supply a minimal set of input files for a pipeline execution; typically sequence reads (FASTQ format) and a reference genome (FASTA format). All reference indices required for an analysis are generated automatically during the pipeline execution. Selected output files and quality reports for downstream interpretation are aggregated into a single directory that may be easily exported and shared. The user may configure to be alerted at the end of a pipeline execution by an email message.

Each of the pipelines comes with a quality report, where available, and with a report summarizing what was done and what are the results. These reports together with the most important output files are automatically copied at the end of a successful analysis into a user-supplied report directory.

The pipelines are processed in one of the three modes of operation, according to the source of sequenced reads: paired-end reads from Illumina sequencers, single-end reads from Nanopore sequencers and Illumina reads processed in single-end mode (Figure 2A). In the case of single-end read sequencing, several tools need to be properly set up to comply with the sequencing technology. For this purpose, an additional key 'platform' has to be supplemented with one of the values {illumina, nanopore}, so that SnakeLines loads appropriate versions of the required tools.

Description of implemented pipelines, as well as instructions for installation, running, and extending the framework, are accessible from the online documentation [https://snakelines.readthedocs.io/](https://snakelines.readthedocs.io/). Moreover, the documentation of a new rule is automatically gathered and deployed when this rule is included in SnakeLines.

# Results

## Implemented pipelines

We implemented several computational pipelines along with examples of input data for the rapid set up of the SnakeLines framework (Table 1). The pipelines reflect our experiences with bioinformatics processing, such as variant calling (Budis et al. 2019; Kubiritova et al. 2019; Budiš et al. 2018), *de novo* assembly (Soltys et al. 2018), metagenomics (Vďačný et al. 2018; Böhmer et al. 2020; Soltys et al. 2020), transcriptomics (Šubr et al. 2020; Misova et al. 2021) and applications in clinical diagnostics (Minarik et al. 2015; Budis et al. 2018; Maronek et al. 2021).

User-supplied sequenced reads are typically preprocessed at first to eliminate sequencing artifacts that may bias downstream analyses. The user may combine several implemented steps: trimming of low-quality ends (Bolger, Lohse, and Usadel 2014), removal of duplicated fragments (Xu et al. 2012), filtering of reads from known hosts or contamination sources (Langmead and Salzberg 2012). Pre-defined numbers of reads from each sample may be selected to avoid variability caused by uneven sequencing depth (Heng Li 2018). Finally, paired-end reads may be merged into singleton fragments based on their sequence overlap (Zhang et al. 2014). The effect of each step may be examined in HTML reports that are automatically generated using standard reporting tools (Andrews 2016; Okonechnikov, Conesa, and García-Alcalde 2016).

Preprocessed reads are passed to a downstream analysis that is chosen by the user according to the biological question to answer. Assembly of reads into contigs (Bankevich et al. 2012), for instance, can be chosen for novel organisms and customized for specifics of bacterial (Wick et al. 2017), metagenomic, transcriptomic, or plasmid-based biological material (Antipov et al. 2016). The quality of assembled contigs may be assessed using summary reports (Gurevich et al. 2013) or visual inspection of *de novo* assembly plots (Wick et al. 2015). Contigs may be further annotated and reviewed in filterable and sortable HTML table with attributes, such as contig length, complexity, GC content of its sequence, homologous sequences in reference databases (Altschul et al. 1990), and sequence similarity with viral genomes (Ren et al. 2017).

Mapping of the reads to a known genomic reference may be also customized according to the specifics of different types of sequenced material; including whole-genome or targeted sequencing (Langmead and Salzberg 2012; Heng Li and Durbin 2009; H. Li 2018), RNA transcripts (Patro et al. 2017), or bisulfite-treated DNA used in the analysis of methylation patterns (Krueger and Andrews 2011). All required indices are built automatically from provided reference sequences. Alternatively, the user may supply a list of accession ids and reference sequences will be automatically downloaded from the Genbank database ([www.ncbi.nlm.nih.gov/genbank/](www.ncbi.nlm.nih.gov/genbank/)), optionally followed by multiple alignments of the sequences (Katoh et al. 2002) visualized in an interactive viewer (Yachdav et al. 2016), or as a phylogenetic tree (Nguyen et al. 2015). SnakeLines also supports variant calling (Lai et al. 2016; McKenna et al. 2010; Poplin et al. 2018; Luo et al., n.d.), as well as thorough classification of DNA fragments originating from a single target gene that are commonly used to study microbial communities (Bengtsson-Palme et al. 2015; Q. Wang et al. 2007; Bolyen et al. 2018) and identification of reads from viral genomes (Tithi et al. 2018). Generated reports are customized according to the type of input material. Besides summary tables and mapping reports for standard genomic material (Okonechnikov, Conesa, and

García-Alcalde 2016), SnakeLines provides specific bar plots and hierarchical pie plots (Ondov, Bergman, and Phillippy 2011) to visually assess the composition of microbial communities. Transcription profiles of RNA-Seq samples can be compared based on principal component analysis (PCA) reduced vectors (Nelli 2015). Transcripts with a significant change in expression between various experimental conditions are identified and summarized in report tables (Robinson, McCarthy, and Smyth 2010).

| Technical aspects | | | | | | | |
|---|---|---|---|---|---|---|---|
| Framework | SnakeLines | MetaMeta | Sequana | VIPER | NGS-pipe | hppRNA | SnakeChunks |
| Language | Python R | Python | Python R | Python R Perl | Python R | Perl R | Python R |
| Installation of framework | Conda | Conda | Conda Pip Singularity | Download source | Download source | Installation script | Download source |
| Installation of dependencies | Automatic on-demand | Conda scripts | Manual | Conda scripts | Conda scripts | Installation script | Manual |
| Configuration | YAML | YAML | YAML | YAML | JSON | Pipeline sources | YAML |
| Graphical interface | - | - | yes | - | - | - | - |
| Implemented pipelines | | | | | | | |
| Assembly | yes | - | yes | - | - | - | - |
| Variant calling | yes | - | yes | - | yes | - | - |
| Metagenomics | yes | yes | - | - | - | - | - |
| Transcriptomics | yes | - | yes | yes | yes | yes | yes |
| Viral identification | yes | yes | - | - | - | - | - |
| Methylation | yes | - | - | - | - | - | - |
| CNV detection | - | - | yes | - | yes | - | - |
| Chip-seq | - | - | - | - | - | - | yes |

Table 1: Comparison of the selected Snakemake-based frameworks for bioinformatics analysis.

## Case study: Variant calling

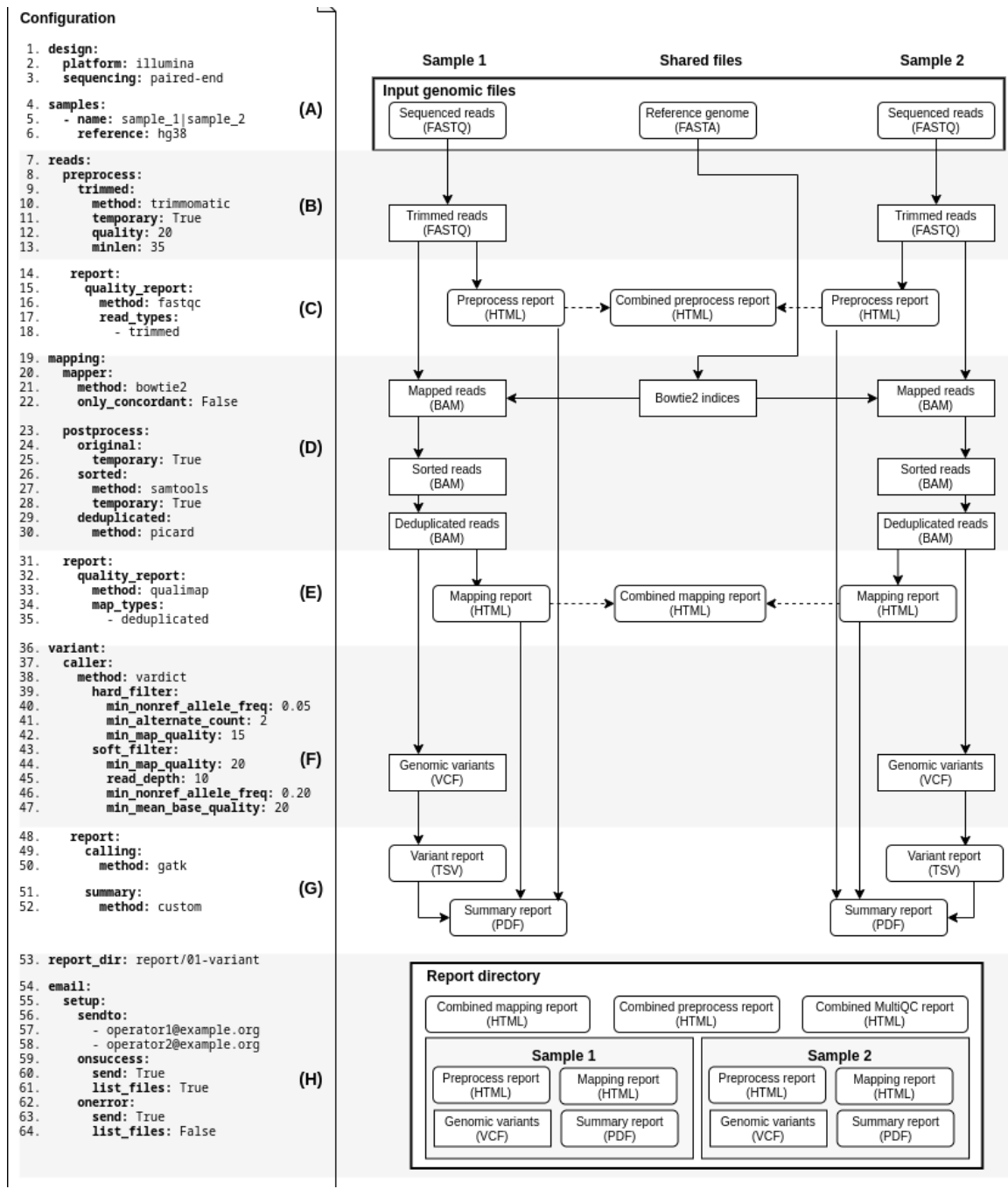

*Figure 2. Basic variant calling pipeline constructed from the user-supplied configuration. (A) SnakeLines runs an analysis on specified FASTQ files (columns Sample 1, Sample 2). Each block of configuration (B)-(G) represents a set of SnakeLines rules that are*

*automatically assembled into computational pipelines using the Snakemake workflow engine. The steps are gradually executed according to the generated workflow. **(H)** Essential output files are copied to the specified directory at the end of the analysis and users are notified by email messages.*

The identification of genomic variation in sequenced reads is a well-studied problem with a wide range of applications (Pabinger et al. 2014). We chose, therefore, a basic variant calling pipeline to describe the fundamental concepts of the SnakeLines framework and how it can be customized through the single configuration file (Figure 2).

At first, the user declares a set of analyzed samples (Sample 1 and Sample 2 in Figure 2A) along with the genomic reference of the sequenced organism (hg38). SnakeLines supports flexible declarations through regular expressions that allow to enumerate names of samples, choose them by pattern, or simply analyze all present samples. The user may also define several sets of samples with different references in a single configuration. SnakeLines checks the presence of the required files in the predefined directories and proceeds with the assembly of the pipeline.

The variant calling pipeline can be divided into three major steps; elimination of sequencing artifacts from sequenced reads (Figure 2B), mapping reads to a reference sequence (Figure 2D), and identification of variation in mapped reads (Figure 2F). Each step is recorded in the configuration to obtain a comprehensive overview of individual analysis steps, the tools used, and their parameters. In such a modular architecture, removing the block corresponding to the variant identification step (Figure 2F, 2G) would effectively reduce the pipeline to the preprocessing and the mapping step. Conversely, adding a block of configuration for other types of analysis, for example, a viral identification would lead to more complex analysis with additional reported files.

Processing steps in standard bioinformatics pipelines have typically well-defined and standardized types of input and output files. For example, mapping transforms sequenced reads in FASTQ format to mapped reads stored in BAM format, while requiring a reference genome in FASTA format. Similarly, variant calling transforms a BAM file into a VCF file. SnakeLines selects a rule that implements required transformation according to the 'method:' attribute that is specified for each processing step of the analysis separately (Figure 2, lines 10, 16, 21, 27, 30, 33, 38, 50, 52). In that manner, changing implementation can be easily done by replacing the value of the attribute. For example the Bowtie2 and the BWA mapper can be readably switched by changing the 'method: bowtie2' to the 'method: bwa' (Figure 2D, line 21). In the case of a novel tool that is not already bundled in the set of SnakeLines rules, the user must at first supply its Snakemake rule template with the same minimal set of inputs and outputs as its alternatives. This rule template must be placed into the same directory as other rules for that specific purpose. Also, the name of the rule template file must match its name, otherwise, SnakeLines would not be able to match the rule to the configuration.

The effect of individual steps may be examined in quality control reports that are stored in human-readable formats, such as HTML or PDF. Again, output reports can be customized, extended, or suppressed in the corresponding configuration blocks (Figure 2C, 2E, 2G). The user may, for example, choose to generate an additional quality report for the original sequenced reads by additional item '- original' in the list of reported read types (Figure 2C, line 17). Reports are aggregated together into combined reports (Ewels et al. 2016) to mitigate laborious inspection of numerous report files generated for each sample separately (Figure 2, dashed arrows).

SnakeLines also supports chaining of transformations for steps that produce the same types of files as types of their inputs. This is particularly useful for the post-process of generated files, for example, to sort a BAM file and then mark duplicated fragments (Figure 2D). Transformations are executed in the order declared in the configuration file. BAM file generated in the last step of the chain (deduplication) is passed to the subsequent variant calling step. To avoid excessive amounts of stored data, the user may choose to remove outputs of these transformations by declaring the 'temporary: True' attribute (Figure 2, lines 11, 25, 28). Marked BAM files would be removed, when no longer needed for further steps of the pipeline.

Bioinformatics pipeline typically generates numerous, mostly auxiliary files, while only a handful of them are typically used for assessing the quality of individual processing steps and interpretation of findings. Essential outputs, quality reports, and files required for reproduction (configuration file and SnakeLines version) are therefore copied automatically to the specified report directory at the end of the analysis (Figure 2H, line 53). Finally, specified users (Figure 2H, line 56) are notified by an email message that can be also sent in case of a failed analysis (Figure 2H, line 62).

# Case study: Benchmarking variant calling pipelines

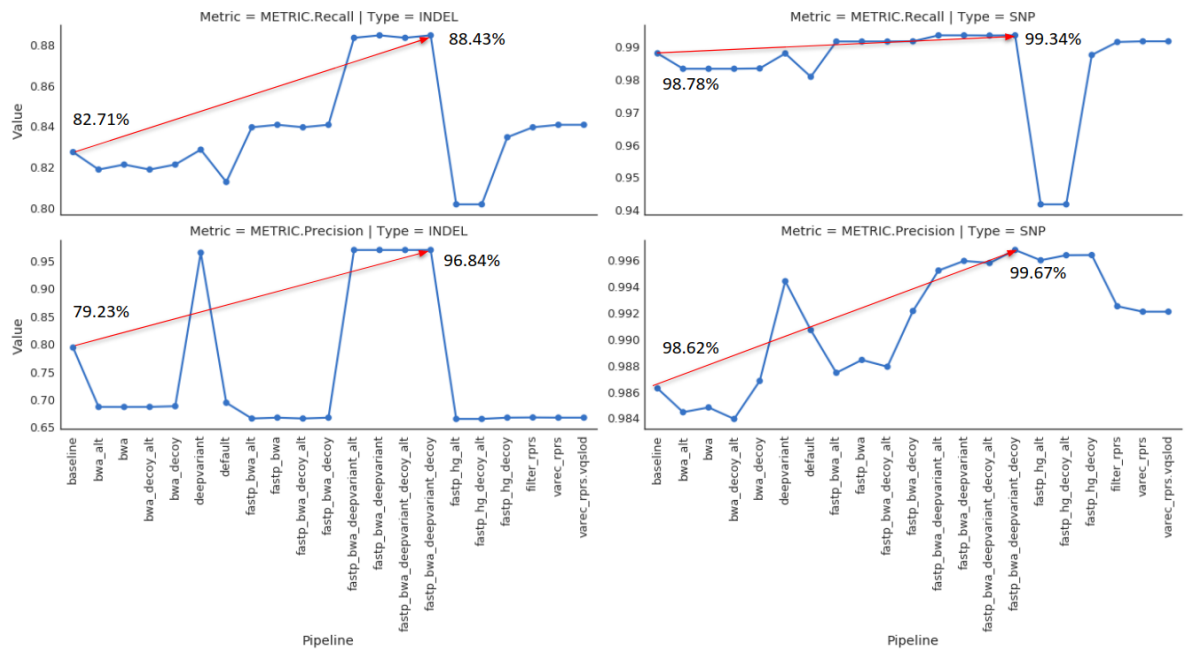

*Figure 3. SnakeLines facilitated the comparison between numerous configurations (only a handful shown for clarity) of variant calling pipeline for single nucleotide variants (right column) and small insertions and deletions (left column). The recall (top row) and precision (bottom row) have been considerably improved (arrows) between the initial (baseline) and the best performing configuration (fastp_bwa_deepvariant_decoy).*

We utilized SnakeLines to find the best performing set of computational steps for variant calling pipeline on the human genome, namely, preprocessing (Figure 2B), mapping (Figure 2D), and variant calling itself (Figure 2F) The configuration-based approach to the set-up and parametrization of the pipeline enabled and greatly eased rapid testing of various pipelines. With the simple adjustments in SnakeLines configuration, we were able to test multiple versions of the human genome (Figure 2A), different computational tools, and parameters. Each configured pipeline ran separately and produced a set of detected variants, along with a report manifest with the precise information of the version of the SnakeLines and the associated configuration, allowing exact reproduction of the results.

The variant calls produced by SnakeLines from each pipeline configuration were compared to the high-confidence reference call set provided by Genome in a Bottle (GIAB) Consortium (Zook et al. 2019) outside of the SnakeLines framework. The comparison was performed according to the best GIAB practices (Krusche et al. 2019) with the tool hap.py (Illumina n.d.). The default pipeline acted as a baseline and comprised of Trimmomatic (Bolger, Lohse, and Usadel 2014), Bowtie2 (Langmead and Salzberg 2012), and Vardict GATK HaplotypeCaller (Langmead and Salzberg 2012; Mulder et al. 2020). We identified the optimal pipeline configuration considering precision and recall metrics on both SNVs and INDELs. This configuration comprised of GRCh38 human reference genome with decoy sequences and without alternative sequences, fastp (Chen et al. 2018) read preprocessor, BWA-MEM mapper (Heng Li and Durbin 2009), and DeepVariant caller (Heng Li and Durbin 2009; Poplin et al. 2018).

## Discussion

SnakeLines is a powerful framework with a set of ready-to-use computational pipelines for several commonly used types of bioinformatics analyses. The configuration has been designed to provide a comprehensive view of all execution steps with emphasis on readability and flexibility in their adjustment and extension. Required bioinformatics tools are installed automatically into isolated virtual environments, which enable the rapid set-up of pipelines on fresh systems and also reproducibility of the analysis across different Unix-based platforms. Due to the powerful features of the underlying Snakemake engine, the execution of pipelines can be easily scaled from a single computer to distributed computational centers.

The SnakeLines framework aims to find the best compromise between easy-to-use graphical workflow managers and the flexibility of command-line-based solutions that require users with a computational background. Although the framework lacks a rich graphical interface, the configuration can be easily handled in any text editor application. Moreover, the basic text format may simplify the set-up of analysis through an external laboratory management system. Laboratory operator has complete control over individual processing steps, as well as implemented tools and their parameters. Although the SnakeLines does not yet provide the broad range of supported bioinformatics pipelines of well-established frameworks, such as Galaxy, the flexible architecture allows to include other tools and processing steps easily, and so have a great potential for further extensions to keep pace with the fast-moving field of nucleic acids analysis.

The presented framework has already become the inherent part of our data processing center (Misova et al. 2021; Strieskova et al. 2019; Böhmer et al. 2020), and so would be further improved and extended with new bioinformatics tools and data analysis pipelines. Although the SnakeLines pipelines have been primarily designed and refined for the paired-end next-generation sequencing reads, the framework allows for readable incorporation of single-end or the third generation sequencing tools. We thus see great potential for a wide use of the framework across other research groups, due to its broad focus, simplicity, and rapid set up of required tools and dependencies.

# Acknowledgments

The presented work was supported by the Slovak Research and Development Agency projects APVV-16-0264 and PP-COVID-20-0051, and the Operational program Integrated Infrastructure within the projects: „Creation of nuclear herds of dairy cattle with a requirement for high health status through the use of genomic selection, innovative biotechnological methods, and optimal management of breeding", NUKLEUS 313011V387, and "Research in the SANET network and possibilities of its further use and development", ITMS code 313011W988, co-financed by the European Regional Development Fund.

# Author Disclosure Statement

JB, WK, MK, RH, ML, DS, AB, FD, JG, JR, TS are the employees of Geneton Ltd. that is a provider of bioinformatics services in Slovakia. All remaining authors have declared no conflicts of interest.